\documentclass[usegraphicx]{mn2e}

\usepackage{times}

\title[Cyclic period changes of a MACHO RRc star]{ A first-overtone RR Lyrae star
with cyclic period changes\thanks{Based
in part on the observations obtained with the 1.3 m Warsaw telescope at Las
Campanas Observatory of the Carnegie Institution of Washington.}}  
\author[A. Derekas et al.]{A. Derekas$^{1,2}$\thanks{E-mail: aliz@phys.unsw.edu.au}, 
L. L. Kiss$^2$, A. Udalski$^3$, T. R. Bedding$^2$, K. Szatm\'ary$^4$ \\  
\\$^1$School of Physics, Department of Astrophysics and Optics, University of 
New South Wales, Sydney, NSW 2052, Australia 
\\$^2$School of Physics, University of Sydney, NSW 2006, Australia 
\\$^3$Warsaw University Observatory, Al. Ujazdowskie 4, PL-00-478 Warsaw, Poland 
\\$^4$Department of  Experimental Physics and Astronomical Observatory, 
University of Szeged, Szeged, D\'om t\'er 9, 6720 Hungary}

\begin{document}

\date{Accepted ... Received ..; in original form ..}


\maketitle

\begin{abstract} A detailed light curve analysis is presented for a
first-overtone RR Lyrae star, MACHO* J050918.712$-$695015.31, based on MACHO
and OGLE-III observations. As a  foreground object of the Large Magellanic
Cloud, it gives an extraordinary  opportunity to study an almost continuous,
12-year long dataset of a relatively  bright ($V\approx15\fm0$) RRc star with
rapid period change. Cyclic period modulation is suggested by the $O-C$
method, where the cycle length is about 8 years. With the available unique
dataset, we could draw strong limits on other light curve changes that may be
associated to the period modulation. We could exclude both multiple periodicity
and amplitude modulation unambiguously. Any theoretical model should reproduce
the observed lack of photometric modulations. Simple arguments are also given
for possible  hydromagnetic effects. \end{abstract}


\section{Introduction}

Period changes in RR Lyrae stars have been an intriguing issue since the 
discovery of this phenomenon (Bailey 1913, Prager 1939). Numerous studies have
been carried out for RR Lyrae stars in globular clusters (e.g., Szeidl 1975,
Wehlau et al. 1992, Rathbun \& Smith 1997, Jurcsik et al. 2001) and in the
galactic field (see a collection of field stars in Firmanyuk 1976, 1982, and a
recent re-analysis of a particular object by Jurcsik et al. 2002).  Stellar
evolution is expected to cause very slow continuous changes in period (Lee
1991, Cox 1998), but the observations tend to reveal abrupt and/or cyclic
changes that are too fast to be explained in this way. The Blazhko effect
(Blazhko 1907), which refers to amplitude modulation in RR~Lyrae stars, does not
yet have a generally accepted explanation (Chadid et al. 2004) and
is also known to be associated with period modulations. To make the overall
picture even more complicated, Papar\'o et al. (1998) found opposite period
changes in the modes of two double-mode RR Lyrae stars in M15. A number of
theoretical studies have attempted to explain period changes in RR Lyrae stars
with various mechanisms, such as mixing events in the semiconvective zone
(Sweigart \& Renzini 1979). Good reviews can be found in Rathbun and Smith
(1997) and Smith (1995, 1997). Understanding these period variations remain an
important unsolved problem in the stellar pulsation theory (Alcock et al.
2000).

Globular clusters that are rich in RR Lyrae variables offer a good opportunity to
study period changes. For example, Rathbun \& Smith (1997) studied the period
changes of RR Lyrae variables in seven globular clusters by collecting all data
in the literature over 50 years. They found that the rates of period changes of
fundamental-mode RR Lyrae (RRab) stars are usually larger than those of
first-overtone (RRc)  and double-mode (RRd) variables. However, they concluded
that current theories did not explain the irregular period changes. More recently,
Jurcsik et al. (2001) presented an analysis of the RR Lyrae variables of $\omega$
Centauri. They studied $O-C$ diagrams of altogether 126 RRab, RRc and RR
Lyrae-like stars and found that the period changes of most RRab were in good
agreement with the evolutionary model predictions, but those of the first-overtone
RRc-type stars showed a much more complex, irregular behaviour (Jurcsik et al.
2001).

Despite the great value of cluster RR Lyrae stars, no one has ever followed a
cluster continuously with the same instrument for many years or decades.  The
essentially random samples over many decades with various instruments do not
provide a homogeneous view of the period changes of RR Lyrae variables. The
situation changed in the early 1990s, when microlensing projects started 
monitoring the sky towards the Large and Small Magellanic Cloud and the Galactic
Bulge, providing continuous photometric observations of millions of stars during
several years. Of these, MACHO and OGLE projects had the best coverage. These data
present a great opportunity for investigating period changes of RR Lyrae stars.
Alcock et al. (2000, 2004) analysed MACHO light curves for more than 1000 RRc
stars and found that 10\% showed strong period changes, whose strength and
patterns excluded simple evolutionary explanations. Meanwhile, Soszy\'nski et al.
(2003) compiled a more complete catalog of $\sim$ 7600 RR Lyrae stars in the LMC
based on the OGLE-II observations. It remains for a future study to combine these
two sources of data for examining period changes over a longer time base.

Although a large amount of data is available for the RR Lyrae stars in the
Large Magellanic Cloud, there is a major drawback with these stars: they are
very faint at the distance of the LMC ($V\approx19\fm0$). Hence, the
signal-to-noise ratio (S/N) of the MACHO observations for the LMC RR Lyrae
stars is very low (Alcock et al. 2000) and neither light curve parameters nor
their changes can be determined accurately. Although OGLE-II RR Lyrae
observations have higher S/N ratio due to better seeing and more advanced
photometric reductions technique, they were carried out over a shorter time
span and with only a single filter.  These problems could be circumvented if we
found suitably bright foreground  stars in the MACHO and OGLE databases. During
an analysis of more than 6000 MACHO variable stars (Derekas et al. in prep.),
we noticed MACHO* J050918.712$-$695015.31 (hereafter M0509$-$69), an RR Lyrae
star with intriguing period changes. At magnitude $V\approx 15\fm0$, period
$P=0.328 {\rm d}$ and amplitude $A_{\rm blue}\approx 0\fm5$, this star offers a
unique opportunity, because of its high S/N ($\sim 80-120$ in the MACHO data
and $\sim 120-130$ in the OGLE-III data) light curve, for analysing the
correlation between period changes and light curve shape variations. This is
expected to tell us something about the physical mechanism causing the
modulations. We note that  M0509$-$69 was also included in the study of Alcock
et al. (1997), who discussed 20 foreground RR Lyrae stars towards the LMC and
noticed the changing period of this star.

The main aim of this work is to demonstrate the period modulation of M0509$-$69,
and show that it is not accompanied by any variations in amplitude or light curve
shape. Our study exploits the homogeneity, the continuity and the 12-year long
time span of the combined MACHO and OGLE-III CCD observations. In Sect. 2 we
describe the data and their basic analysis. We show several pieces of  evidence
for  pure period modulation of M0509$-$69 in Sect. 3. Finally, we briefly discuss
possible mechanisms in Sect. 4. 

\section{Initial data analysis}

Our attention was drawn to this star during an on-going study of MACHO variable
stars, that are classified as eclipsing binaries in the on-line MACHO Variable 
Star Catalog. The MACHO observations were obtained between June 1992 and January
2000, with the 50 inch telescope at Mt. Stromlo Observatory using a two-channel
system (Hart et al. 1996). Two-colour data in the specifically designed MACHO blue
($B_{\rm M}$) and MACHO red ($R_{\rm M}$) bands are publicly available at the
MACHO website\footnote{\tt http://wwwmacho.mcmaster.ca/}. Stars have been
classified by an automatic software and we downloaded individually every target,
6837 in total. However, it became obvious very quickly that the classification was
not perfect, as large fraction of ``eclipsing binaries'' was found to be Cepheids,
RR Lyrae stars or long-period variables. We therefore decided to re-classify all
stars and during this procedure we noted the peculiar light curve of M0509$-$69.
Whereas the period and amplitude unambiguously place this star among the
first-overtone RR Lyraes (Poretti 2001), the light curve showed enormous phase
modulations.

\begin{figure*}  
\includegraphics[width=17cm]{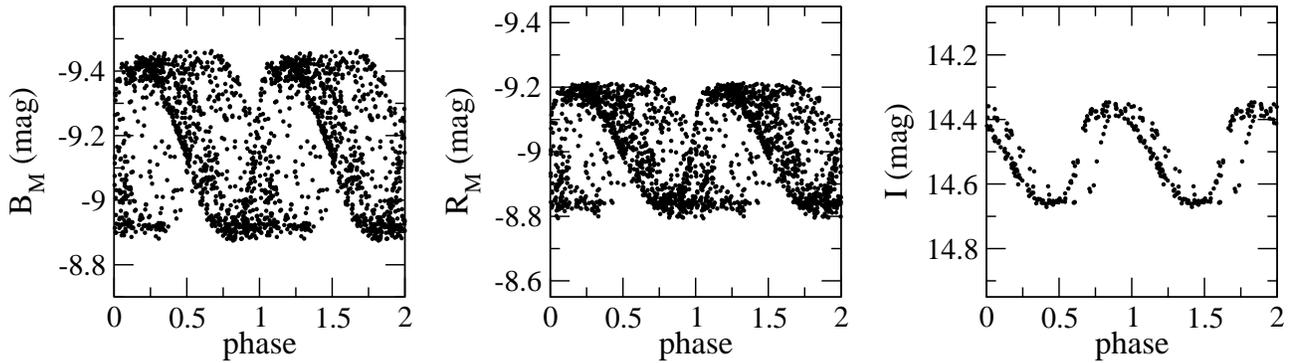}   
\caption{The phase diagram for M0509$-$69 in three colours: MACHO blue, MACHO red
and OGLE.}
\label{phase}  
\end{figure*}

In light of the strange behavior of M0509$-$69 in the MACHO data, we decided to
supplement them with OGLE observations. Unfortunately, this object is located
outside the fields covered during the second phase of the OGLE project (Udalski et
al. 1997) so that OGLE-II photometry from 1997--2000 is not available. However,
this field was covered in the OGLE-III phase, which started in June 2001, and the
star is continuously monitored up to now. Thus the gap between the MACHO and
OGLE-III observations is only 621 days. The OGLE observations were obtained with
the 1.3 m Warsaw telescope located at Las Campanas Observatory, Chile, which is
operated by the Carnegie Institution of Washington, equipped with $8192\times
8192$ pixel mosaic camera. Photometry was derived with the standard OGLE data
pipeline based on DIA image subtraction technique (Udalski 2003). Observations
were done through $I$-band filter.

As the first step of the analysis of M0509$-$69, we searched for the period using
the  Phase Dispersion Minimization method (Stellingwerf 1978), then we refined
that period using the String Length method (Lafler \& Kinman 1965, Clarke 2002).
The phase diagram with the finally adopted period (Sect.\ 3.2) is plotted in Fig.\
\ref{phase}. Both the phase diagram and random checks of short subsets suggested a
strong period modulation, so we therefore determined epochs of maximum light to
construct the classical $O-C$ diagram. To this, we divided the whole dataset into
50-day long subsets, which did not show any noticeable phase shift. Then we phased
each subset with the period given by the String Length method (0.32806 days). When
there were too few points within 50 days, then we used 100-day long subsets. The
best of these phase diagrams was fitted with a fourth-order Fourier-polynomial.
This polynomial was used as a master curve -- allowing both vertical and
horizontal shifts to fit every other phase diagram. The same procedure was applied
for the OGLE-III data. From the resulting phase shifts, the period and  the epochs
of each  phase diagram, we calculated  the times of maximum light, listed in
Table\ \ref{maxtimes}. Typical uncertainties are about 0.001--0.0015 days.

Finally, we note that downloadable MACHO data are in Modified Julian Date, so 
that we had to add 0.5 days to all epochs of maximum. 

\begin{table}
\begin{center}
\caption{Times of maximum for M0509$-$69.}
\label{maxtimes}
\begin{tabular}{|rcrc|}
\hline
$HJD_{\rm max}$ & filter  & $HJD_{\rm max}$ & filter \\
\hline
2448885.1603 & ${\rm R_M}$ & 2449937.4734 & ${\rm B_M}$\\
48885.1712 & ${\rm B_M}$ & 50037.2174 & ${\rm B_M}$\\
48948.1480 & ${\rm B_M}$ & 50039.1748 & ${\rm R_M}$\\
48950.1030 & ${\rm R_M}$ & 50139.2403 & ${\rm R_M}$\\
48997.0187 & ${\rm B_M}$ & 50139.2513 & ${\rm B_M}$\\
49001.2712 & ${\rm R_M}$ & 50241.9485 & ${\rm B_M}$\\
49045.2343 & ${\rm B_M}$ & 50243.9034 & ${\rm R_M}$\\
49050.1431 & ${\rm R_M}$ & 50342.3490 & ${\rm B_M}$\\
49095.0899 & ${\rm B_M}$ & 50349.2265 & ${\rm R_M}$\\
49099.9961 & ${\rm R_M}$ & 50436.1742 & ${\rm R_M}$\\
49146.9115 & ${\rm B_M}$ & 50446.3549 & ${\rm B_M}$\\
49152.4765 & ${\rm R_M}$ & 50543.1468 & ${\rm B_M}$\\
49197.4241 & ${\rm B_M}$ & 50547.0712 & ${\rm R_M}$\\
49200.3659 & ${\rm R_M}$ & 50649.4351 & ${\rm R_M}$\\
49253.1684 & ${\rm R_M}$ & 50653.3818 & ${\rm B_M}$\\
49253.1796 & ${\rm B_M}$ & 50749.4961 & ${\rm R_M}$\\
49292.5371 & ${\rm R_M}$ & 50749.5064 & ${\rm B_M}$\\
49303.3733 & ${\rm B_M}$ & 50850.2163 & ${\rm R_M}$\\
49397.1685 & ${\rm R_M}$ & 50850.2268 & ${\rm B_M}$\\
49397.1801 & ${\rm B_M}$ & 50955.2051 & ${\rm B_M}$\\
49447.0279 & ${\rm R_M}$ & 51034.5776 & ${\rm R_M}$\\
49452.2878 & ${\rm B_M}$ & 51074.6031 & ${\rm B_M}$\\
49497.2208 & ${\rm B_M}$ & 51120.1905 & ${\rm R_M}$\\
49497.5391 & ${\rm R_M}$ & 51181.5408 & ${\rm B_M}$\\
49547.4005 & ${\rm R_M}$ & 51290.1142 & ${\rm B_M}$\\
49551.3507 & ${\rm B_M}$ & 51317.9810 & ${\rm R_M}$\\
49597.2600 & ${\rm R_M}$ & 51394.4042 & ${\rm B_M}$\\
49599.2387 & ${\rm B_M}$ & 51432.4460 & ${\rm R_M}$\\
49648.4325 & ${\rm R_M}$ & 51502.3273 & ${\rm B_M}$\\
49652.3801 & ${\rm B_M}$ & 51516.0922 & ${\rm R_M}$\\
49697.3256 & ${\rm B_M}$ & 52227.1610 & I\\
49699.2805 & ${\rm R_M}$ & 52333.7701 & I\\
49749.1449 & ${\rm R_M}$ & 52578.8074 & I\\
49752.1087 & ${\rm B_M}$ & 52667.7100 & I\\
49834.1273 & ${\rm B_M}$ & 52895.0690 & I\\
49836.0847 & ${\rm R_M}$ & 52995.7926 & I\\
49936.4784 & ${\rm R_M}$ & &\\
\hline
\end{tabular}
\end{center}
\end{table}

\section{Results}

\subsection{Multiple periodicity}

\begin{figure} 
\includegraphics[width=8.5cm]{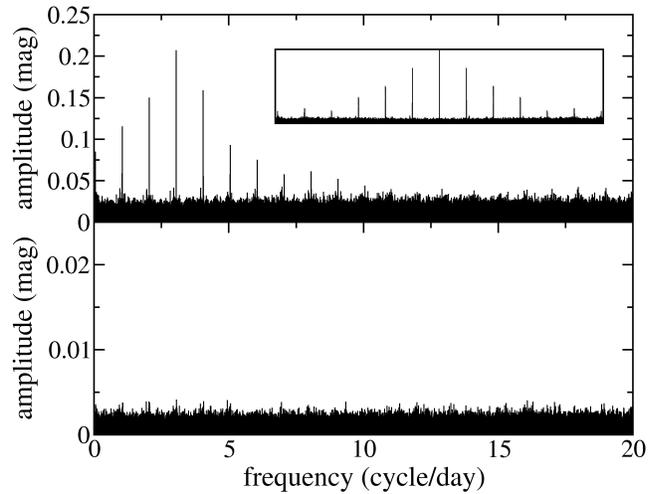} 
\caption{The Fourier spectra of the complete blue colour dataset ({\it{top 
panel}}) and  the residuals, after removal of the main period, harmonics and 
aliases ({\it{bottom panel}}). The insert shows the window function on same 
frequency scale.} 
\label{fourier} 
\end{figure}

We examined the possibility of multiple periodicity in the MACHO blue-band light
curve of M0509$-$69. We chose this set because of the higher amplitude, thus
better S/N. The frequency analysis was carried out by standard Fourier-analysis
using Period98 of  Sperl (1998). The calculated amplitude spectrum is shown in the
top panel of Fig.\ \ref{fourier}. The main peak at $f_0=3.048222 d^{-1}$ is in
very good agreement with the result by the String Length method  ($3.048223
d^{-1}$). Usual prewhitening steps resulted in the same behaviour as the one found
for RR1-PC-type stars by Alcock et al. (2000): the prewhitened frequency spectrum
contained significant remnant power very close to the main component. As was shown
in Alcock et al. (2000), this can be attributed to long-term period and/or
amplitude change. Since the procedure of determining epochs of maximum proved the
stability of the amplitude (i.e. every phase diagram could be fitted very well
with the master curve), we checked for the presence of multiple periodicity in two
different ways. Firstly, we did regular prewhitening in the best three 50-day long
subsets to check the residual frequency spectra for any significant peak unrelated
to the pulsational frequency. For that reason, we fitted and subtracted not only
$f_0$, but $2f_0$, $3f_0$ and $4f_0$ from the subsets. None of the residual
spectra contained a significant peak other than $5f_0$, after which we could not
find anything else. 

Secondly, we subtracted individually fitted Fourier-polynomials from every subset
in form of   \begin{equation} m=A_{0}+\sum_{i=1}^{4} A_{i} \cdot \sin\left( 2 \pi
ift + \phi_{i} \right) ,  \end{equation}  where the amplitudes and phases were
fitted with Period 98. After subtracting each polynomial, we re-united the
residual light curves and calculated the frequency spectrum of the whole set. As
expected from the asymmetric shape of the mean light curve, we detected only 
$5f_0$ and $6f_0$ with amplitudes of $0\fm0064$ and $0\fm0045$. After subtracting
these two frequencies, the S/N ratios (Breger et al. 1993) of the next three peaks
were about 4, close to the limit of significance. We could identify these
frequencies as close remnants of the primary peak and its harmonics or daily
aliases. Finally, we arrived at pure white noise (bottom panel in Fig.\
\ref{fourier}), confirming the monoperiodic nature of M0509$-$69. Similar results
were given by the red data too, but their S/N is lower than that of the blue data,
which we preferred in our analysis.

\subsection{The O$-$C diagram}

Having excluded multiple periodicity, the $O-C$ diagram was made from the times of
maxima in Table\ \ref{maxtimes}.Initially, we used the period given by the String
Length method, with the following ephemeris: $HJD_{\rm max}=2449197.4241 + 0.32806
\times E$. It showed very clear  cyclic period changes without colour dependence
(i.e. the blue and the red $O-C$ points drew the same curve).

\begin{figure}
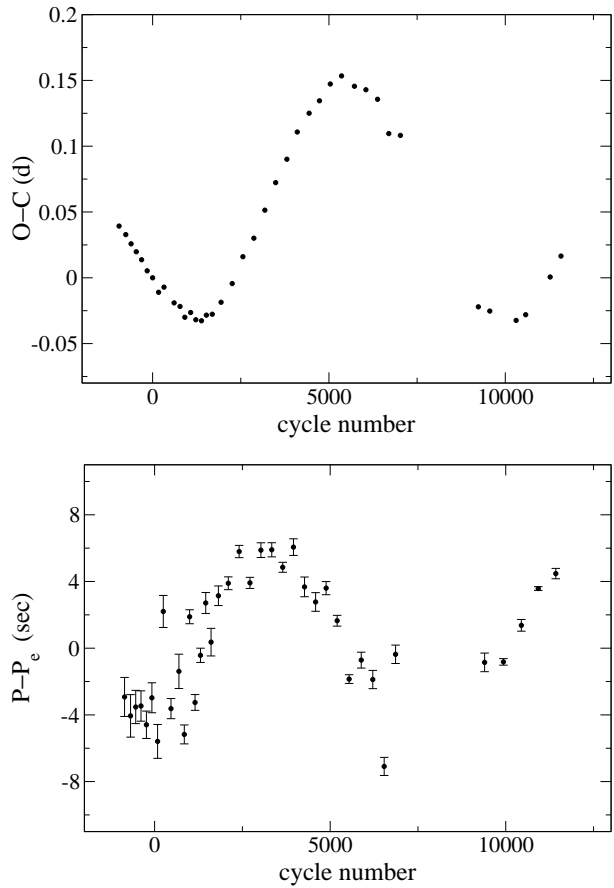

\includegraphics[width=8cm]{ME614_3.eps}
\vskip3mm
\includegraphics[width=8cm]{ME614_4.eps}
\caption{{\it{Top panel}}: the $O-C$ diagram of M0509$-$69 from $B_{\rm M}$ and $I$
data. {\it{Bottom panel}}: the "instantaneous" period. The error bars are the $3 
\sigma$ errors of the fits.}
\label{o-c}
\end{figure}

There is a well-known problem with the $O-C$ method: although it is a very
effective tool to detect tiny period changes, the shape of the diagram strongly
depends on the period used in the calculation. Moreover, it can be discontinuous
after long periods, so that we analysed the $O-C$ diagram using the continuous
method suggested by Kalimeris et al. (1994). In this approach, both the period and
its rate of change are continuous functions of time. Rather than using various
assumptions on period change (e.g. assuming constant change of period and thus
fitting parabola to the $O-C$; or fitting straight lines to selected parts of the
$O-C$), we determined the ``instantaneous'' periods from every neighbouring pair
of subsets, simply as the local derivative of the $O-C$ plot plus the period used
in the ephemeris ($P_e$). The difference between the ``instantaneous'' period and
$P_e$ is plotted in the bottom panel of Fig.\ \ref{o-c}. As can be seen here, the
period change is enormous: the difference between the longest and the shortest
period is about 12 seconds; in other words, the relative period change is
$\pm2\cdot 10^{-4}$!

The use of the ``instantaneous'' period has a further advantage: this way we can
combine MACHO and OGLE-III data without involving systematic phase shift between
the $B_{\rm M}$ and $I$ bands. We checked several foreground RR Lyraes with
simultaneous MACHO and OGLE-II observations (Zebrun et al. 2001) and found that
I-band maxima are usually within 0.01$-$0.03 in phase relative to $ B_{\rm
M}$-band maxima. With the ``instantaneous'' period, even this small effect is
eliminated.

Finally, we corrected the period by assuming that the time-span of the combined
MACHO+OGLE-III data is long enough to average out the period modulation. In
other words, we searched that period which resulted in zero mean period
difference in the bottom panel of Fig.\ \ref{o-c}. The resulting ephemeris is
the following:
\begin{equation} 
HJD_{\rm max}=2449196.9241 + 0.328038(2) \times E, 
\end{equation} 
and the finally adopted $O-C$ diagram in the top panel of Fig.\ \ref{o-c} was
also calculated with this ephemeris.

\subsection{Amplitude modulation}

Some RR Lyrae stars show cycle-to-cycle changes in the shape of their light
curves, which can be caused either by double-mode pulsation (in RRd stars) or 
the still-puzzling Blazhko-effect. Since double-mode pulsation was ruled out in
Sect. 3.1, what remains to be checked is the possibility of amplitude modulation.

\begin{figure} 
\includegraphics[width=8.5cm]{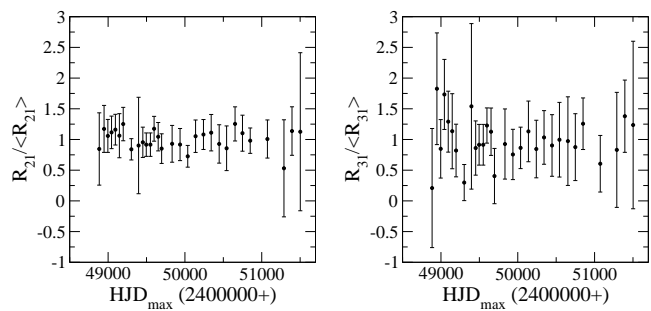} 
\caption{The relative variations of $R_{21}$ and $R_{31}$.}
\label{fiterror} 
\end{figure}

Although the master curve fitted very well all subsets, we  wanted to examine
quantitatively the possibility of light curve shape change. Therefore, we fitted 
fourth-order Fourier-polynomials to each subset using equation (1) at the primary
frequency and its harmonics. Then we measured the light curve shapes with the
Fourier parameters (Simon and Teays 1982), of which we show particular results for
$R_{21}=A_{2}/A_{1} \ {\rm and} \ R_{31}=A_{3}/A_{1}$ (these are the most accurate
parameters).  Fig.\ \ref{fiterror} shows the relative variations of $R_{21}$ and
$R_{31}$. The error bars are quite large (because the light curve is not too
asymmetric, so that higher harmonics are fairly weak), but it is evident that
there were no measurable shape changes.

\begin{figure*}
\includegraphics[width=17cm]{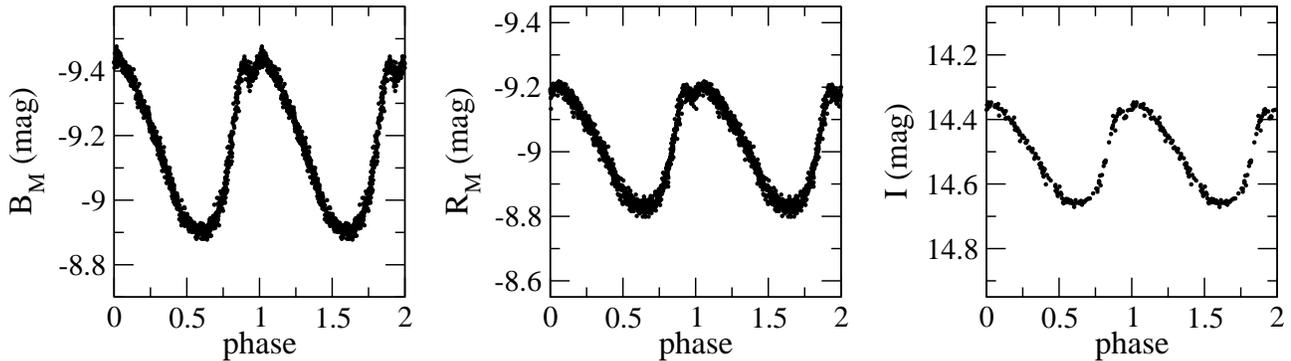}
\caption{The phase diagrams after eliminating the phase shifts between all subsets.}
\label{subphase}
\end{figure*}

Since individual subsets rarely contained enough data points for determining
accurate Fourier-parameters, we also examined the correlation between the period
change  and the light curve shape using five subsets around the minimum and five
around maximum of the ``instantaneous'' period. Then we corrected these phase
diagrams for any phase shifts to form high-quality, continuous phase diagrams
representing the two extrema of the period. Then we fitted sixth-order  Fourier
polynomials to both phase diagrams and the resulting parameters are shown in
Table\ \ref{fit}. As can be seen, none of the Fourier-coefficients differs
significantly for the two cases. This means that we could not detect any light
curve change associated with the period variation. This conclusion is also
supported by the phase diagram of the whole MACHO and OGLE-III datasets (Fig.\
\ref{subphase}), where we shifted every subset to match the master curve in phase.
The well-defined shapes show no evidence for any additional variation in light
curve shape.

\begin{table}
\begin{center}
\caption{The coefficients of the sixth-order Fourier polynomials around the 
minimum and the maximum of the period.}
\label{fit}
\begin{tabular}{|ccc|}
\hline
parameter & values at $P_{\rm min}$  & values at $P_{\rm max}$\\
\hline
$a_{0}$ & --9.165 $\pm$ 0.003 & --9.168 $\pm$ 0.001\\
$a_{1}$ & --0.278 $\pm$ 0.005 & --0.270 $\pm$ 0.005\\
$a_{2}$ & --0.046 $\pm$ 0.004 & --0.044 $\pm$ 0.005\\
$a_{3}$ & 0.023 $\pm$ 0.005 & 0.022 $\pm$ 0.005\\
$a_{4}$ & 0.024 $\pm$ 0.004 & 0.016 $\pm$ 0.005\\
$a_{5}$ & --0.017 $\pm$ 0.005 & --0.010 $\pm$ 0.005\\
$a_{6}$ & --0.015 $\pm$ 0.004 & --0.008 $\pm$ 0.005\\
$\phi_{1}$ & 1.249 $\pm$ 0.016 & 1.251 $\pm$ 0.018\\
$\phi_{2}$ & --3.780 $\pm$ 0.104 & --3.639 $\pm$ 0.112\\
$\phi_{3}$ & 0.602 $\pm$ 0.182 & 0.492 $\pm$ 0.228\\
$\phi_{4}$ & 2.007 $\pm$ 0.199 & 2.303 $\pm$ 0.312\\
$\phi_{5}$ & --0.181 $\pm$ 0.256 & --0.313 $\pm$ 0.476\\
$\phi_{6}$ & 0.332 $\pm$ 0.324 & 0.155 $\pm$ 0.580\\
\hline
\end{tabular}
\end{center}
\end{table}

\section{Discussion}

It is obvious from Fig.\ \ref{o-c} that the period change seems to be cyclic, 
which implies there must be some (quasi)periodic mechanism affecting the
pulsation. The shape of the curve in the bottom panel of Fig.\ \ref{o-c} is
largely independent of the period used for the $O-C$ diagram (provided that it
is close to the real mean period), so that the observed change cannot be
explained by some numerical artifact caused by a wrong period in the ephemeris.

As a first possibility, we consider light-time effect (LITE) due to orbital motion
in a binary system. Although it was suggested for several RR Lyrae stars that
their cyclic $O-C$ diagrams were caused by LITE, so far this was supported by
spectroscopic observations only for TU UMa (Wade et al. 1999). For M0509$-$69, the
half-amplitude of the $O-C$ in Fig.\ \ref{o-c} is A(O$-$C)=0.09 d, while the time
separation of the two minimum is 2900 d. Assuming that $A(O-C)=\frac{a\sin 
i}{c}$, the estimated semi-major axis is about $15.5/\sin i$ AU. Given $P_{\rm
orb}$=2900 d $\approx$ 7.95 yr; the minimum mass of the system would be almost
$60M_{\odot}$.

Although there are models that predict the existence of detached black hole
binaries with $M_{BH}=10-20$ M$_\odot$ and $M_{sec}=0.3-1$ M$_\odot$ (see, e.g.,
Table 1 in Podsiadlowski et al. 2003), the secondary components in these
theoretical systems are not expected to follow the evolution of ``normal'' stars
with the same masses. This is because they are descendants of more massive
progenitor stars that transferred most of their masses to the black hole primaries
(but models also predict that they can evolve through the classical instability
strip, even becoming ``RR Lyrae-like'' variables). In principle, we cannot exclude
this possibility until it is checked by spectroscopy (any sign of orbital 
motion should be relatively easily detectable spectroscopically even at 
$V=15$ mag), although $60M_{\odot}$ 
seems extraordinary large. Also, the high occurrence rate of strong period 
changes in RR Lyrae stars argues against this exotic explanation.

Stellar evolution can be safely excluded, since both the strength and the cyclic
nature of the period change exceed theoretical expectations by orders of
magnitudes (Smith 1995).

Another interesting possibility is related to hypothetic hydromagnetism.  Stothers
(1980) argued that RR Lyrae stars may be considered as fair analogues of the Sun.
He proposed that radius changes may be driven by some sort of magnetic activity,
which in turn would cause observable period changes. The correlation between the
frequency changes and the magnetic variations was confirmed by Howe et al. (2002),
who performed a detailed study on the solar cycle frequency shifts in
global $p$-modes of the Sun. They found that the latitudinal distributions of the
frequency shifts shows close temporal and spatial correlation with the unsigned
surface magnetic flux.

In our case we can calculate the radius change of M0509$-$69 from the 
period-density relation, i.e. $P \sqrt{\rho}={\rm Q}$. Substituting the values
determined from the bottom panel in Fig.\ \ref{o-c}, it follows: 
\begin{equation} \frac{\Delta R}{R}=\frac{2}{3} \cdot \frac{\Delta P}{P} 
\approx \frac{2}{3} \cdot
\frac{6}{28342} \approx 1.4 \cdot 10^{-4}.    
\end{equation}

\noindent Note that this equation is an approximation that omits a term that
contains the magnetic energy (see Stothers 1980).

Very recently, No\"el (2004) presented an analysis of 13 years of apparent radius
measurements of the Sun. Interestingly, he found a correlation between the solar
radius variation and the sunspot numbers, related to the solar magnetic activity.
Based on the data in his fig. 5, the relative change of the solar radius was about
$\Delta R/R \approx 4\cdot 10^{-4}$. Assuming that RR Lyrae stars can show
solar-like magnetic activity cycles, these two similar relative radius changes
seem to support the hypothetic hydromagnetic period change.

We know little about the presence of magnetic fields in RR~Lyrae stars.
Most recently, Chadid et al. (2004) presented high-precision longitudinal magnetic
field measurements of RR~Lyrae itself taken over 4 years. Their results provided no
evidence for a strong magnetic field in the photosphere of this star, which was often
postulated in various explanations for the Blazhko effect (see, e.g., 
Jurcsik et al. 2002). If the period change in M0509--69 is indeed caused by 
solar-type magnetic activity, this field might be quite difficult to detect
observationally. In fact, the only Blazhko-related phenomena that have been detected
so far seem to be small line-profile variations (possibly related to non-linear
pulsations; Chadid et al. 1999, Chadid 2000) and even these are only detectable 
with high S/N high-resolution spectra.

Although the effects of hydromagnetism on pulsating stars are not well 
understood, we conclude that the Stothers-mechanism can perhaps explain the pure
period modulation of M0509$-$69. However, new theoretical investigations are
necessary to interpret this observed behaviour. Detailed models should also
address the absence of any light curve shape change and multiple periodicity.

\section*{Acknowledgments} 

This work has been supported by the FKFP Grant 0010/2001, OTKA Grant \#T042509 
and the Australian Research Council. AD is supported by the International
Postgraduate Research Scholarship (IPRS) programme of the Australian Department of
Education, Science and Training. Partial support to the OGLE
project was provided with the Polish KBN grant 2P03D02124, NSF grant AST-0204908
and NASA grant NAG5-12212. The NASA ADS Abstract Service was used to access data
and references. This paper utilizes public domain data obtained by the MACHO
Project, jointly  funded by the US Department of Energy through the University of
California,  Lawrence Livermore National Laboratory under contract No.
W-7405-Eng-48, by the  National Science Foundation through the Center for Particle
Astrophysics of the  University of California under cooperative agreement
AST-8809616, and by the  Mount Stromlo and Siding Spring Observatory, part of the
Australian National  University.

\end{document}